\documentclass[aps,prc,preprint,tightenlines,floatfix,groupedaddress,
nofootinbib,showpacs,preprintnumbers,amsmath,amssymb,superscriptaddress]
{revtex4}
\usepackage{graphicx}
\usepackage{dcolumn}
\usepackage{mathrsfs}
\usepackage{bm}

\begin{document}

\title{Do we understand the incompressibility of neutron-rich matter?}
\author{J. Piekarewicz\footnote{\tt e-mail: jpiekarewicz@fsu.edu}}
\affiliation{Department of Physics, Florida State University,
             Tallahassee, FL {\sl 32306}}
\date{\today}

\begin{abstract}
The {\sl ``breathing mode''} of neutron-rich nuclei is our window into
the incompressibility of neutron-rich matter. After much confusion on
the interpretation of the experimental data, consistency was finally
reached between different models that predicted both the distribution
of isoscalar monopole strength in finite nuclei and the compression
modulus of infinite matter. However, a very recent experiment on the
Tin isotopes at the {\sl Research Center for Nuclear Physics}~(RCNP)
in Japan has again muddled the waters. Self-consistent models that
were successful in reproducing the energy of the giant monopole
resonance (GMR) in nuclei with various nucleon asymmetries (such as
${}^{90}$Zr, ${}^{144}$Sm, and ${}^{208}$Pb) overestimate the GMR
energies in the Tin isotopes. As important, the discrepancy between
theory and experiment appears to grow with neutron excess. This is
particularly problematic as models artificially tuned to reproduce the
rapid softening of the GMR in the Tin isotopes become inconsistent
with the behavior of dilute neutron matter. Thus, we regard the 
question of {\sl ``why is Tin so soft?''} as an important
{\sl open problem in nuclear structure}.
\end{abstract}
\maketitle 


\section{Incompressibility of Symmetric Nuclear Matter}
\label{KSNM}

The incompressibility coefficient of infinite nuclear matter---also 
known as the compression modulus---is a fundamental parameter 
of the equation of state (EOS). The compression modulus controls 
small density fluctuations around the equilibrium point, thereby 
providing the first glimpse into the \textsl{``stiffness"} of the EOS.
Whereas existing ground-state observables ({\sl e.g.,} masses and
charge radii) have accurately constrained the saturation point of 
{\sl symmetric} nuclear matter (at a baryon density of 
$\rho_{{}_{0}}\!\simeq\!0.15~{\rm fm}^{-3}$ and a binding energy per 
nucleon of $\varepsilon_{0}\!\simeq\!-16$~MeV), the extraction of 
the compression modulus ($K_{0}$) is significantly more complicated 
as it requires to probe the response of the nuclear ground state.  It 
is widely accepted that the nuclear compressional modes---particularly 
the quintessential {\sl ``breathing mode''} or isoscalar {\sl Giant
Monopole Resonance} (GMR)---provide the cleanest, most direct route 
to the nuclear incompressibility~\cite{Blaizot:1980tw, Blaizot:1995}.

Earlier attempts at extracting the compression modulus of symmetric
nuclear matter relied primarily on the distribution of isoscalar
monopole strength in ${}^{208}$Pb---a heavy, doubly-magic nucleus 
with a well developed monopole 
peak~\cite{Youngblood:1977,Youngblood:1981}. Although such 
measurements have existed for some time, the field has enjoyed a 
renaissance due to new and improved experimental facilities and 
techniques. Indeed, an improved $\alpha$-scattering experiment
found the position of the giant monopole resonance in ${}^{208}$Pb 
at $E_{\rm GMR}\!=\!14.17\pm0.28$~MeV~\cite{Youngblood:1999}.  
As this measurement was combined with the distribution of monopole 
strength in other nuclei---and compared with microscopic
calculations---a value of the incompressibility coefficient in the
range of $K_{0}\!=$220-240~MeV was extracted.

As the experimental program reached a high level of maturity and
sophistication, the same strict standards were demanded from the
theoretical program. Indeed, calculations of nuclear compressional
modes based on consistent {\sl Mean-Field plus Random Phase
Approximation} (RPA) approaches became routine. Moreover, such
consistent models---without any recourse to semi-empirical mass
formulas---were able to simultaneously predict the incompressibility
of infinite nuclear matter as well as the distribution of isoscalar
monopole strength in finite nuclei~\cite{Pearson:1991, Shlomo:1993zz}.
However, as the experimental story was coming to an end, the
theoretical picture remained unclear. On the one hand, nonrelativistic
calculations that reproduced the distribution of isoscalar-monopole
strength in ${}^{208}$Pb predicted a nuclear incompressibility
coefficient in the $K_{0}\!=$210-230~MeV
range~\cite{Colo:1992,Blaizot:1995,Hamamoto:1997b}.  On the other
hand, relativistic models that succeeded in reproducing a large body
of nuclear observables---including the GMR in ${}^{208}$Pb---suggested
the significant larger value of
$K_{0}\!\simeq\!270$~MeV~\cite{Lalazissis:1996rd,Vretenar:2001hs}.

The solution to this puzzle was originally proposed by
Piekarewicz~\cite{Piekarewicz:2002jd,Piekarewicz:2003br} and has since
been confirmed by several other
groups~\cite{Vretenar:2003qm,Agrawal:2003xb,Colo:2004mj}.  The
solution is based on the realization that the GMR in ${}^{208}$Pb does
not constrain the compression modulus of {\sl symmetric} nuclear
matter but rather the one of {\sl neutron-rich} matter.  In
particular, it was concluded that the compression modulus of a neutron
rich system having the same neutron excess as ${}^{208}$Pb is {\sl
lower} than the compression modulus of symmetric nuclear matter. This
could explain how models with significantly different
incompressibility coefficients $K_{0}$ may still reproduce the GMR in
${}^{208}$Pb~\cite{Piekarewicz:2002jd,Piekarewicz:2003br}.  As a
result, accurately-calibrated theoretical models were built to
reproduce simultaneously the distribution of isoscalar-monopole
strength in both ${}^{90}$Zr and ${}^{208}$Pb---nuclei with a well
developed monopole peak yet significantly different nucleon
asymmetries~\cite{Todd-Rutel:2005fa,Agrawal:2005ix}.  Since then, the
large difference in the predicted value of $K_{0}$ between
nonrelativistic and relativistic models has been reconciled and a
``consensus'' has been reached that places the value of the
incompressibility coefficient of {\sl symmetric} nuclear matter at
$K_{0}\!=\!240\!\pm\!10$~MeV~\cite{Agrawal:2003xb,Colo:2004mj,
Todd-Rutel:2005fa,Lalazissis:2005de,Garg:2006vc}.

\section{Incompressibility of Neutron-Rich Matter}
\label{KNRM}

To summarize some of the above findings it is convenient to discuss in
detail the incompressibility coefficient of infinite, neutron-rich
matter. On very general grounds---indeed, in a
model-independent way---the incompressibility coefficient of
neutron-rich matter may be written as
\begin{equation}
  K_{0}(\alpha)=K_{0}+K_{\tau}\alpha^{2}+{\cal O}(\alpha^{4}) \;,
  \label{K0Alpha}
\end{equation}
where $\alpha\!\equiv\!(N\!-\!Z)/A$ is the nucleon asymmetry.  From
this expression it is immediately evident that the GMR in ${}^{208}$Pb
(with a neutron excess of $\alpha\!=\!0.21$) should be sensitive to a
linear combination of the incompressibility coefficient of symmetric
nuclear matter $K_{0}$ and $K_{\tau}$---a quantity that determines the
evolution of the incompressibility coefficient with neutron excess.
Note that $K_{\tau}$ plays the same role in determining the
incompressibility coefficient as the symmetry energy at saturation
density (a quantity often denoted by $J$ or $a_{4}$) plays in
determining the energy-per-nucleon of asymmetric matter.

To compute the incompressibility coefficient of neutron-rich matter
one proceeds exactly as in the case of symmetric nuclear matter.
First, one determines the equilibrium point at a fixed value of
$\alpha$ and then extracts $K_{0}(\alpha)$ from computing the
curvature at the minimum. Having done so for various values of
$\alpha$, one can extract $K_{\tau}$ from a simple fit to
Eq.~(\ref{K0Alpha})~\cite{Piekarewicz:2008nh}.  An alternative
procedure that is highly accurate and significantly more illuminating
starts from the equation of state of neutron-rich matter parametrized
in terms of several bulk parameters defined at normal saturation
density. Starting from such a parametrization, it becomes a simple
exercise to compute the equilibrium point (as a function of $\alpha$)
and the corresponding curvature at the minimum. In particular, one
obtains the following closed-form expression for
$K_{\tau}$~\cite{Piekarewicz:2008nh}:
\begin{equation}
 K_{\tau}=K_{\rm sym}-6L-\frac{Q_{0}}{K_{0}}L\;.
 \label{KTau}
\end{equation}
where $L$ and $K_{\rm sym}$ represent the slope and curvature of the
symmetry energy at saturation density~\cite{Piekarewicz:2008nh}.
Although often neglected, note that $K_{\tau}$ also depends on the
third derivative of the EOS of symmetric nuclear matter $Q_{0}$ (a
quantity often referred to as the {\sl ``skewness''} parameter). Quite
generally, as the infinite nuclear system becomes neutron rich, the
saturation density moves to lower densities, the binding energy
weakens, and the nuclear incompressibility {\sl
softens}~\cite{Piekarewicz:2008nh}. It is important to stress the
dominant role of the {\sl symmetry pressure} $L$ on these conclusions
and in particular on Eq.~(\ref{KTau}) (because of the large
coefficient in front of it). We note that the symmetry pressure
$L$---a quantity that strongly influences the neutron-skin thickness
of heavy nuclei---is directly proportional to the pressure of {\sl
pure neutron matter}, namely,
\begin{equation}
  P_{nm} = \frac{1}{3}\rho_{{}_{0}}L\;.
 \label{PZero}
\end {equation} 
This connection is important as significant theoretical progress has
been made in constraining the equation of state of low-density neutron 
matter. We will draw heavily on this connection in what follows.

\section{Measuring the Breathing Mode of the Tin Isotopes}
\label{KNRM}

The realization that the distribution of monopole strength in heavy
nuclei is sensitive to the density dependence of the symmetry energy
motivated a recent experimental study of the GMR along the isotopic
chain in Tin~\cite{Garg:2006vc,Li:2007bp}. This important experiment
was carried out at the {\sl Research Center for Nuclear
Physics}~(RCNP) in Osaka, Japan. Such an experiment probed the
incompressibility of {\sl asymmetric} nuclear matter by measuring the
distribution of isoscalar strength in a chain of isotopes ranging from
${}^{112}$Sn (with $\alpha\!=\!0.11$) to ${}^{124}$Sn (with
$\alpha\!=\!0.19$).  Because of the sensitivity of $K_{\tau}$ to the
pressure of pure neutron matter [see Eqs.~(\ref{KTau})
and~(\ref{PZero})], this experiment represents an attractive hadronic
complement to the {\sl purely electroweak} Parity Radius Experiment
(PREx) at the Jefferson Laboratory that aims to measure the neutron
radius of $^{208}$Pb accurately and model independently via
parity-violating electron
scattering~\cite{Horowitz:1999fk,Michaels:2005}.  We note that such a
fundamental measurement will have far-reaching implications in areas
as diverse as nuclear structure~\cite{Todd:2003xs}, heavy-ion
collisions~\cite{Danielewicz:2002pu,Tsang:2004,Chen:2004si,Li:2005sr,
Shetty:2005qp}, atomic parity
violation~\cite{Todd:2003xs,Sil:2005tg,Piekarewicz:2006vp} and nuclear
astrophysics~\cite{Horowitz:2000xj,Horowitz:2001ya,Horowitz:2002mb,
Carriere:2002bx,Buras:2003sn,Lattimer:2004pg,Lattimer:2006xb,Steiner:2004fi}.

\begin{figure}[ht]
\vspace{0.50in}
\includegraphics[width=5.0in,angle=0]{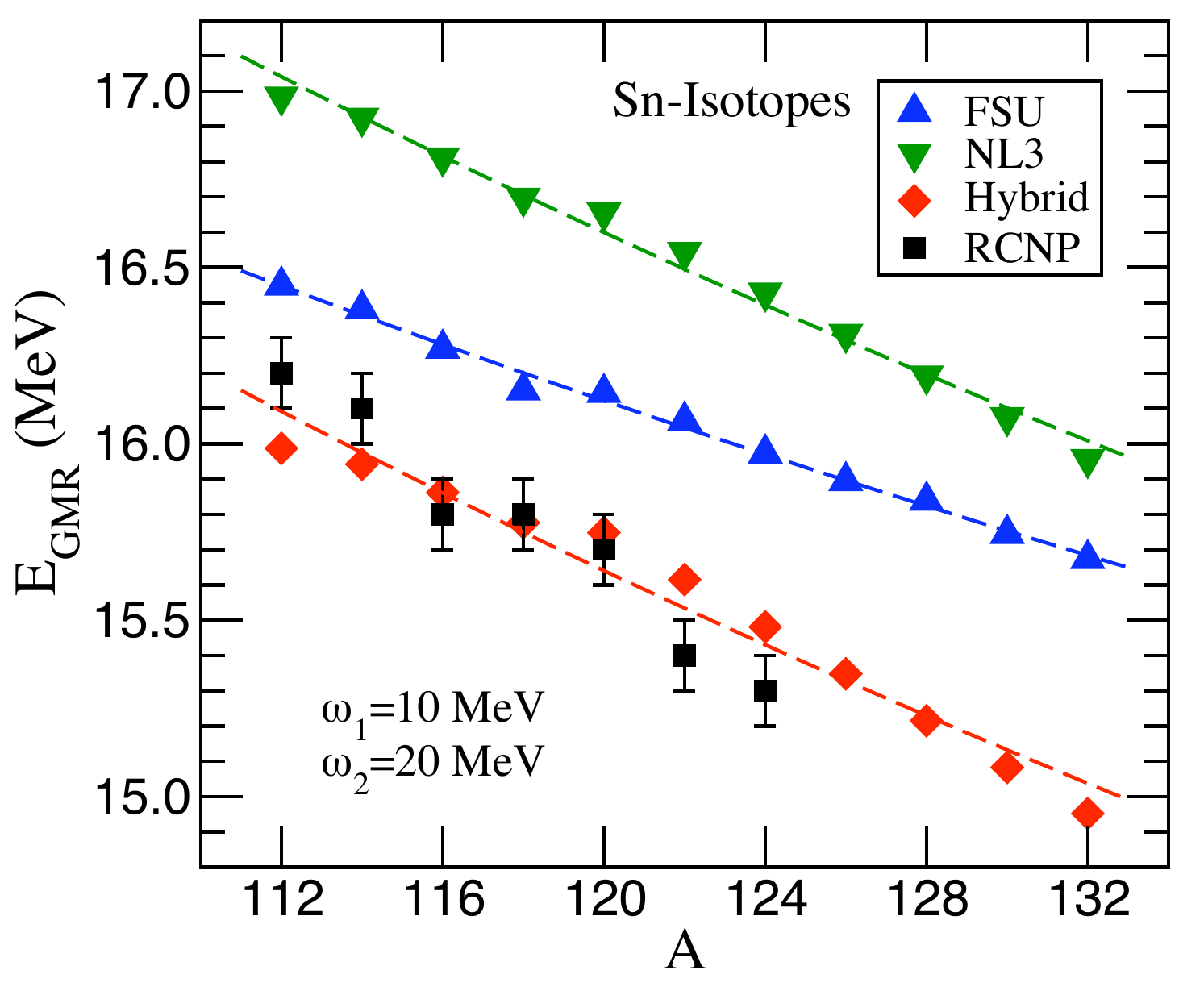}
\caption{(Color online) Comparison between the GMR 
         centroid energies ($m_{1}/m_{0}$) of all neutron-even 
         ${}^{112}$Sn-${}^{124}$Sn isotopes extracted from 
         experiment~\cite{Garg:2006vc,Li:2007bp}
         (black solid squares) and the theoretical predictions 
         from the FSUGold (blue up-triangles), NL3 (green 
         down-triangles), and Hybrid (red diamonds) models. 
         The corresponding dashed lines were obtained from
         a fit to the centroid energies of the form
         $E_{\rm GMR}=E_{0}A^{-\lambda}$ [see Eq.~(\ref{EGMRFit})].}
\label{Fig1}
\end{figure}

Shortly after the completion of the RCNP experiment a serious
discrepancy was revealed: accurately calibrated models that reproduce
the GMR in ${}^{90}$Zr, ${}^{144}$Sm, and ${}^{208}$Pb overestimate
the distribution of isoscalar monopole strength in the Tin
isotopes~\cite{Piekarewicz:2007us,Sagawa:2007sp,
Tselyaev:2009}. Moreover, the discrepancy between theory and
experiment appears to grow with neutron excess $\alpha$, suggesting
that the models significantly {\sl underestimate} the value of
$|K_{\tau}|$. We have colloquially referred to this problem as {\sl
``why is Tin so soft?''}. To illustrate this situation we display in
Fig.~\ref{Fig1} a comparison between the experimental centroid
energies~\cite{Garg:2006vc,Li:2007bp} for the neutron-even
${}^{112}$Sn-${}^{124}$Sn isotopes and three theoretical calculations
that have been extended up to the doubly magic nucleus ${}^{132}$Sn. All
theoretical predictions were generated using a consistent RPA
approach.  That is, the linear response of the system was computed
using the {\sl same} interaction employed to generate the mean-field
ground state. A detailed description of this approach may be found in
Refs.~\cite{Piekarewicz:2000nm,Piekarewicz:2001nm}.

\begin{figure}[ht]
\vspace{0.50in}
\includegraphics[width=5.0in,angle=0]{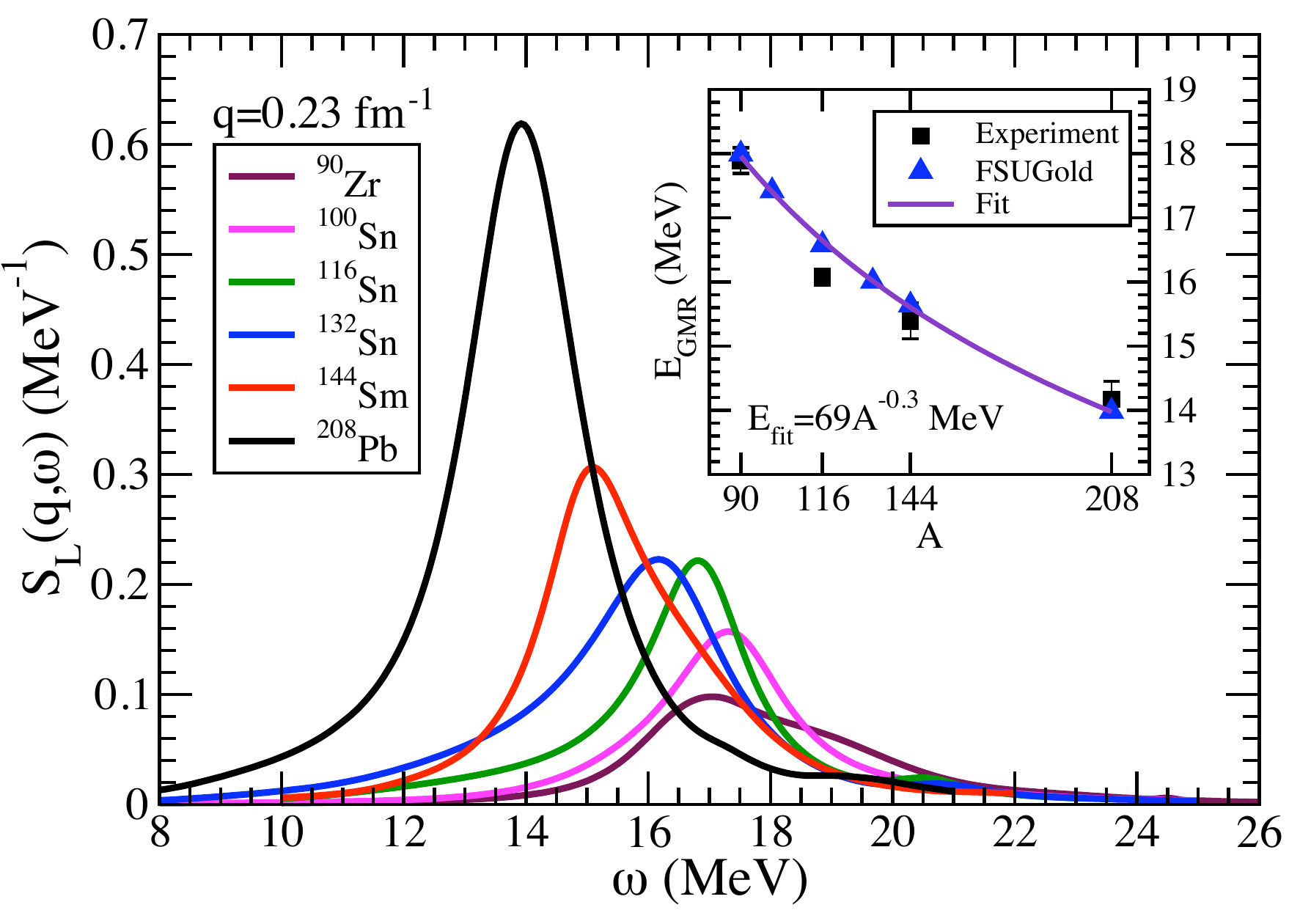}
\caption{(color online) Distribution of isoscalar monopole 
          strength predicted by the FSUGold model of 
          Ref.~\cite{Todd-Rutel:2005fa}. The inset includes 
          a comparison against the experimental centroid 
          energies reported in Ref.~\cite{Youngblood:1999}, 
          with the solid line providing the best fit to the 
          theoretical predictions.}
\label{Fig2}
\end{figure}

The results depicted with the blue triangles were generated using the
accurately calibrated FSUGold
parametrization~\cite{Todd-Rutel:2005fa}.  This relativistic model is
characterized by a soft behavior for both symmetric nuclear matter and
the symmetry energy. (Note that the terms {\sl ``soft''} and {\sl
``stiff''} refer to whether the energy increases slowly or rapidly
with density.) Such a soft behavior is reflected in the relatively
small values of $K_{0}$, $L$, and $|K_{\tau}|$ listed in
Table~\ref{Table1}. Clearly, the model overestimates the experimental
data (black squares). Moreover, the discrepancy increases with neutron
number: from about $0.2$~MeV for ${}^{112}$Sn to $0.7$~MeV for
${}^{124}$Sn. Such a serious discrepancy is particularly troublesome
given that the same FSUGold model successfully reproduces the centroid
energy of the GMR in ${}^{90}$Zr, ${}^{144}$Sm, and ${}^{208}$Pb, as
shown in the inset of Fig.~\ref{Fig2}.  Figure~\ref{Fig2} also displays the
distribution of isoscalar monopole strength from which the centroid
energies depicted in the inset were computed (as the ratio of the first
to the zeroth moment). In addition to the four nuclei---${}^{90}$Zr,
${}^{116}$Sn, ${}^{144}$Sm, and ${}^{208}$Pb---of
Ref.~\cite{Youngblood:1999}, we display the distribution of monopole
strength for the doubly-magic nuclei ${}^{100}$Sn and ${}^{132}$Sn. We
note that the GMR predictions for all six nuclei fall nicely within
the {\sl ``liquid-drop''} inspired curve $E_{\rm
GMR}\!\simeq\!69A^{-1/3}$~\cite{BohrII:1998,Harakeh:2001}.  Moreover,
these predictions reproduce the experimentally extracted GMR energies
~\cite{Youngblood:1999}---except for the case of
${}^{116}$Sn. Figures~\ref{Fig1} and~\ref{Fig2} capture the essence of
the problem of {\sl why is Tin so soft?}

\begin{table}
\begin{tabular}{|l||c|c|c|c||c|c|c|c|}
 \hline
 Model & $\rho_{0}$ & $\varepsilon_{0}$
       & $K_{0}$ & $Q_{0}$ & $J$ & $L(P_{nm})$
       & $K_{\rm sym}$  & $K_{\tau}$ \\
 \hline
 \hline
 FSU &  0.148   & $-$16.30 & 230.0 & $-$523.4 & 32.59 &
 60.5(3.0) & $-$51.3  & $-$276.8   \\
 NL3 &  0.148   & $-$16.24 & 271.5 & $+$204.2 & 37.29 & 
 118.2(5.8) & $+$100.9 & $-$697.4  \\
 Hybrid  &  0.148   & $-$16.24 & 230.0 & $-$71.5  & 37.30 & 
 118.6(5.9) & $+$110.9  &  $-$563.9 \\
\hline
\end{tabular}
\caption{Bulk parameters characterizing the behavior of
              neutron-rich matter around saturation density 
              $\rho_{{}_{0}}$. The quantities
              $\varepsilon_{0}$, $K_{0}$, and $Q_{0}$ represent
              the binding energy per nucleon, incompressibility
              coefficient, and third derivative (or ``skewness''
              coefficient) of symmetric nuclear matter at
              $\rho_{{}_{0}}$.
              Similarly, $J$, $L$, and $K_{\rm sym}$ represent 
              the energy, slope, and curvature of the symmetry
              energy at saturation density. All quantities are in 
              MeV, except for $\rho_{{}_{0}}$ which is given in
              ${\rm fm}^{-3}$ and the pressure of pure neutron 
              matter at saturation density ($P_{nm}$) which is 
              given in MeV/fm${}^{3}$. A detailed explanation
              of all these quantities may be found in 
              Ref.~\cite{Piekarewicz:2008nh}.}
\label{Table1}
\end{table}

Also shown in Figure~\ref{Fig1} are the predictions from the NL3
parameter set~\cite{Lalazissis:1996rd,Lalazissis:1999}. The NL3 set
has been remarkably successful in reproducing a myriad of nuclear
ground-state properties (such as masses, charge-radii, and
deformations) throughout the periodic table.  Although it now seems
likely that the stiff behavior predicted by NL3 may be unrealistic, at
the time of its inception most of the information in favor of a softer
equation of state was unavailable. Thus, although NL3 may reproduce
the GMR in ${}^{208}$Pb, the data on the Tin isotopes provides ample
evidence that such a stiff behavior is inconsistent with data. Note,
however, that although NL3 significantly overestimates the GMR
energies in the Tin isotopes, the softening of the incompressibility
coefficient (namely, its dependence with $A$) appears consistent with
data. That is, Figure~\ref{Fig1} suggests that whereas NL3 has too
large a value of $K_{0}$, its value for $K_{\tau}$ may be consistent
with experiment (see Eq.~(\ref{K0Alpha}) and Table~\ref{Table1}).

Motivated by the above facts, we have built a {\sl ``Hybrid ''} model
with a low incompressibility coefficient and a stiff symmetry
energy~\cite{Piekarewicz:2008nh}. However, unlike the NL3 and FSUGold
parametrizations, the Hybrid model was not accurately
calibrated. Thus, the Hybrid model should be simply regarded as a {\sl
``test''} model that illustrates how surprisingly soft are the
experimental GMR energies of the Tin isotopes relative to the
theoretical predictions. We observe in Fig.~\ref{Fig1} that the Hybrid
model yields a significant improvement in the description of the
experimental data. Indeed, the predictions of the Hybrid model fall
within $0.1$~MeV of the experimental data for the full isotopic
chain. Note that relative to FSUGold, the improved description
provided by the Hybrid model is entirely due to its large negative
asymmetric term $K_{\tau}$, as they shared the same value of $K_{0}$
(see Eq.~(\ref{K0Alpha}) and Table~\ref{Table1}). Indeed, we can
capture the $A$ dependence of the $E_{\rm GMR}$ predicted by all the
models with a liquid-drop inspired formula of the form 
$E_{\rm GMR}\!\simeq\!E_{0}A^{-\lambda}$. We obtain,
\begin{equation}
 E_{\rm GMR} = \begin{cases} 
     64.5 A^{-0.29} & \text{for FSU,} \\
   102.6 A^{-0.38} & \text{for NL3,} \\
   112.7 A^{-0.41} & \text{for Hybrid.} 
\end{cases}
\label{EGMRFit}
\end{equation}
The Hybrid model suggests a falloff with $A$ that is significantly
faster than the $\lambda\!=\!-1/3$ value predicted by a liquid-drop
description~\cite{BohrII:1998,Harakeh:2001}. And although the
improvement in the case of the Tin isotopes is significant and
unquestionable, an important problem remains: the Hybrid model
\textsl{underestimates} the GMR centroid energy in ${}^{208}$Pb by
almost 1 MeV~\cite{Tselyaev:2009,Piekarewicz:2008nh}. This 
suggests that the rapid softening with neutron excess predicted 
by the Hybrid model may be unrealistic.

\section{Why is Tin so soft?}

So why is Tin so soft and why does it become even softer as the
nucleon asymmetry increases? Are we any closer to the answer now than
we were then~\cite{Garg:2006vc,Li:2007bp} ? Unfortunately not! As we
elaborate below, we will assume that the experimental extraction of
the GMR energies is without error---although the large discrepancy
between the RCNP~\cite{Garg:2006vc,Li:2007bp} and the Texas
A\&M~\cite{Lui:2004wm} results should be resolved.

To date, only two plausible scenarios have been advanced to explain
why accurately-calibrated models that reproduce the GMR energies in
${}^{90}$Zr, ${}^{144}$Sm, and ${}^{208}$Pb fail to do so for the Tin
isotopes. One of them is encapsulated in the Hybrid model discussed
above~\cite{Piekarewicz:2008nh} whereas the other one suggests that
pairing correlations are responsible for the softening of the monopole
response~\cite{Li:2008hx,Khan:2009xq}.  As discussed earlier, the
Hybrid model is based on an effective interaction that generates a
soft EOS for symmetric nuclear matter (a small $K_{0}$) but a stiff
symmetry energy (a large $|K_{\tau}|$). Any such model should generate
soft monopole excitations for symmetric ($N\!=\!Z)$ nuclei and
significantly softer ones for the neutron-rich isotopes (see
Fig.~\ref{Fig1}). Unfortunately, the Hybrid model---and others like
it~\cite{Tselyaev:2009}---can only reproduce the GMR energies in Tin
at the expense of significantly underestimating the GMR energy in
${}^{208}$Pb.  In the case of pairing correlations, the explanation is
based on the conjecture that a superfluid---such as the open-shell Tin
isotopes---may be easier to compress than a normal
fluid~\cite{Khan:2009xq}. Whereas the validity of this statement is
both interesting and presently unknown, recent Quasiparticle RPA
(QRPA) calculations seem to support the conjecture---at least in
part~\cite {Li:2008hx,Khan:2009xq}. ``At least in part'' because
although pairing correlations yield an appreciable softening for the
lighter isotopes (from ${}^{112}$Sn to ${}^{120}$Sn), the discrepancy
for the heavier ones (${}^{122}$Sn and ${}^{124}$Sn) remains
large~\cite{Li:2008hx,Khan:2009xq}.  This indicates that pairing
correlations can not account for the observed softening of the GMR
energies with nucleon asymmetry. To make matters worse, we now argue
that the rapid softening displayed by the experimental GMR energies in
the Tin isotopes may be even harder to explain as one incorporates the
physics of dilute neutron matter.

\begin{figure}[ht]
\vspace{0.50in}
\includegraphics[width=5.0in,angle=0]{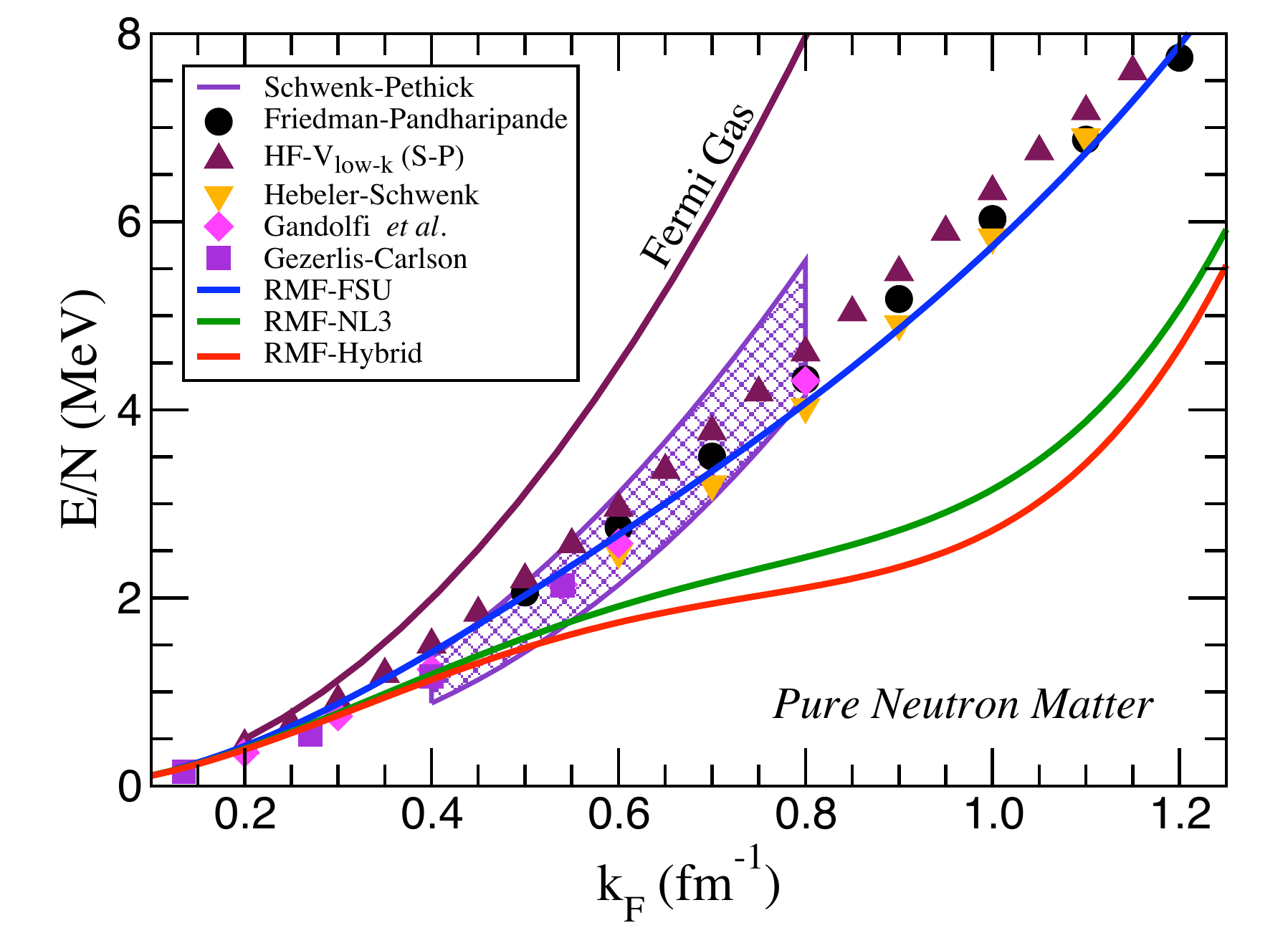}
\caption{(color online) Equation of state of pure neutron 
         matter as predicted from a variety of microscopic 
         models~\cite{Friedman:1981qw,Schwenk:2005ka,
         Hebeler:2009iv,Gandolfi:2008id, Gezerlis:2009iw}. Also 
         shown are predictions from the relativistic mean-field 
         models FSUGold~\cite{Todd-Rutel:2005fa}, 
         NL3~\cite{Lalazissis:1996rd,Lalazissis:1999},
         and Hybrid~\cite{Piekarewicz:2008nh}.}
\label{Fig3}
\end{figure}

\section{Low-Density Neutron Matter}
\label{LDNM}

By building on the universal behavior of dilute Fermi gases with an
infinite scattering length~\cite{Baker:1999dg,
Heiselberg:2000bm,Carlson:2003,Nishida:2006br}, significant progress
has been made in constraining the equation of state of pure neutron
matter. One of the biggest challenges in understanding dilute neutron
matter arises from the non-negligible effective range of the
neutron-neutron ($nn$) interaction ($r_{\rm e}\!=\!+2.7$~fm) which,
although significantly smaller than the scattering length
($|a|\!=\!18.5$~fm), induces important corrections to the universal
behavior at relatively low densities ($k_{\rm F}\!\simeq\!1/r_{\rm
e}\!\lesssim\!0.4~{\rm fm}^{-1}$). To date, a host of models using
different $nn$ interactions and a variety of many-body techniques have
been employed to compute the EOS of dilute neutron matter (see
Fig.~\ref{Fig3}).  These models range from the venerated equation of
state of Friedman and Pandharipande~\cite{Friedman:1981qw}, to those
based on modern effective field theory
approaches~\cite{Schwenk:2005ka, Hebeler:2009iv}, to those using
sophisticated {\sl ``ab-initio''} Monte Carlo
techniques~\cite{Gandolfi:2008id, Gezerlis:2009iw}, to name just a few
(for a more comprehensive list see
Ref.~\cite{Gezerlis:2009iw}). Remarkably, all these {\sl microscopic}
models are fairly consistent with each other.

Also shown in Fig.~\ref{Fig3} are the predictions from three
relativistic mean-field models whose parameters have been fitted
directly to various properties of finite nuclei. By directly fitting to the
experimental data, the parameters of these models encode physics (such
as few- and many-body correlations) that goes beyond a simple
single-particle picture.  In this regard, the underlying parameters of
the model may have, at best, a tenuous connection to those appearing
in microscopic descriptions of the nucleon-nucleon ($NN$)
interaction. As a result, if the mean-field models are not
sufficiently constrained by experimental data, they can predict
behavior that is inconsistent with microscopic approaches---and
with nature. This is clearly the case for the NL3 and Hybrid models
displayed in Fig.~\ref{Fig3}, and for most of the older relativistic
parametrizations. (Note that the energy of pure neutron matter is to a
very good approximation equal to the energy of symmetric nuclear
matter plus the symmetry energy).  Given that existing ground-state
observables do not place stringent constraints on the isovector $NN$
interaction, most relativistic mean-field models predict a stiff
symmetry energy, namely, one that increased rapidly with density for
$\rho\!\gtrsim\!0.1~{\rm fm}^{-3}$ (see the large values of $L$ and
$K_{\rm sym}$ listed in Table~\ref{Table1}).  In contrast, the FSUGold
parametrization incorporates collective modes directly into the fit,
including GMR energies for both ${}^{90}$Zr (with $\alpha\!=\!0.11)$
and ${}^{208}$Pb (with $\alpha\!=\!0.21)$. As a result, a
significantly softer symmetry energy ensues. And although there is 
no {\sl a-priori} guarantee, it is gratifying to observe that the
softening of the symmetry energy displayed by the FSUGold model 
is consistent with the EOS predicted by the various microscopic
approaches (see Fig.~\ref{Fig3}). We suggest that the equation of
state of pure neutron matter provides a powerful constraint that
should be routinely and explicitly incorporated into future
determinations of energy density functionals.

So what is the connection between the largely model-independent  
equation of state of dilute neutron matter and the incompressibility 
coefficient of neutron-rich matter? To appreciate this connection we
first combine Eqs.~(\ref{K0Alpha}) and~(\ref{KTau}) to write the 
incompressibility coefficient of asymmetric matter as
\begin{equation}
  K_{0}(\alpha)=K_{0}+
 \left(K_{\rm sym}-6L-\frac{Q_{0}}{K_{0}}L\right)\alpha^{2}+
 {\cal O}(\alpha^{4}) \;.
 \label{K0Alpha2}
\end{equation}
Then, as the energy of pure neutron matter is to an excellent
approximation equal to the sum of the energy of symmetric 
matter plus the symmetry energy, the EOS of pure neutron 
matter around saturation density may be expressed in 
terms of a conveniently defined dimensionless parameter 
$x=(\rho-\rho_{{}_{0}})/{3\rho_{{}_{0}}}$ and the same bulk
parameters appearing in Eq.~(\ref{K0Alpha2}). That is,
\begin{equation}
  E_{nm}/N =
 (\varepsilon_{{}_{0}}+J)+Lx + 
 \frac{1}{2}(K_{0}+K_{\rm sym})x^{2}+
 \frac{1}{6}(Q_{0}+Q_{\rm sym})x^{3} + \ldots
 \label{EvsX}
\end{equation}
Thus, the evolution of the incompressibility coefficient with nucleon
asymmetry may be parametrized in terms of two bulk parameters of the
EOS of symmetric nuclear matter ($K_{0}$ and $Q_{0}$) and the slope
and curvature of either the symmetry energy ($L$ and $K_{\rm sym}$) or
of pure neutron matter ($L_{nm}\!\equiv\!L$ and
$K_{nm}\!=\!K_{0}+K_{\rm sym}$).  That is,
\begin{equation}
  K_{\tau} = K_{\rm sym}-6L-\frac{Q_{0}}{K_{0}}L = 
                 K_{nm}-6L_{nm}-\frac{K_{0}^2+Q_{0}L_{nm}}{K_{0}}\;.
 \label{KTau2}
\end{equation}
This establishes a strong correlation between $K_{\tau}$ and the
density dependence of the EOS of pure neutron matter (through $L_{nm}$
and $K_{nm}$). In most accurately-calibrated models the dominant
contribution to $K_{\tau}$ comes from the slope of the symmetry energy
$L\!=\!L_{nm}$~\cite{Chen:2007ih,Chen:2009wv}. Indeed, in the three
relativistic mean-field models considered here the dominant term
($6L$) accounts for at least 75\% of the value of $K_{\tau}$.  Given
this fact, we believe that values as large as
$|K_{\tau}|\!\simeq\!550$~MeV---as seem to be suggested by
experiment~\cite{Garg:2006vc,Li:2007bp}---are inconsistent with
the behavior of dilute neutron matter.

\section{Conclusions}
\label{Conclusions}

The present contribution centered around the recently measured
distribution of isoscalar monopole strength in the Tin
isotopes~\cite{Garg:2006vc,Li:2007bp}.  This critical experiment
suggests a significant softening of the GMR energies that is
unexplained by existing theoretical models. Before the publication of
the experimental data in 2007~\cite{Garg:2006vc,Li:2007bp}, there was
strong evidence in support of a value of the incompressibility
coefficient of {\sl symmetric} nuclear matter around
$K_{0}\!=\!240\!\pm\!10$~MeV.  However, the measurement on the Tin
isotopes has forced us to pause and re-examine our
models. Particularly confusing is the fact that some of these
accurately-calibrated models are successful in reproducing the GMR
energies in ${}^{90}$Zr, ${}^{144}$Sm, and ${}^{208}$Pb.  {\sl So why
is Tin so soft and why does it become even softer with an increase in
the neutron excess?} One possible explanation relies on the open-shell
structure of the Tin isotopes and its assumed superfluid
character~\cite{Li:2008hx,Khan:2009xq}. Although this approach has met
with some success, the impact of pairing correlation on the heavy Tin
isotopes (${}^{122}$Sn to ${}^{124}$Sn) is modest so the discrepancy
remains.  Another approach---encapsulated in the Hybrid model
discussed above and introduced in
Ref.~\cite{Piekarewicz:2008nh}---adopts a small value for the
incompressibility coefficient of symmetric matter and a large value
for the slope of the symmetry energy.  This {\sl ``soft-stiff''}
combination is fairly successful in describing the rapid softening of
the GMR energies in the Tin isotopes. However, the same model
significantly {\sl underestimates} the GMR energy in
${}^{208}$Pb. Moreover, the EOS of neutron matter generated by the
Hybrid model---and essential to reproduce the rapid softening of the
GMR in the Tin isotopes---is inconsistent with microscopic models that
based their predictions in the universality of dilute Fermi gases.

In conclusion, the distribution of isoscalar monopole strength in the
Tin isotopes remains an important open problem in nuclear structure.
As one attempts to solve this difficult problem, one must remember
that the challenge is not solely to describe the distribution of
monopole strength in the Tin isotopes, but rather, to do so while 
simultaneously describing a host of ground-state observables, 
collective modes, and the equation of state of low-density neutron 
matter.

\begin{acknowledgments}
 This work was supported in part by a grant from the U.S. 
 Department of Energy DE-FD05-92ER40750. 
\end{acknowledgments}

\vfill\eject
\bibliography{Incompressibility.bbl}

\begin{thebibliography}{62}
\expandafter\ifx\csname natexlab\endcsname\relax\def\natexlab#1{#1}\fi
\expandafter\ifx\csname bibnamefont\endcsname\relax
  \def\bibnamefont#1{#1}\fi
\expandafter\ifx\csname bibfnamefont\endcsname\relax
  \def\bibfnamefont#1{#1}\fi
\expandafter\ifx\csname citenamefont\endcsname\relax
  \def\citenamefont#1{#1}\fi
\expandafter\ifx\csname url\endcsname\relax
  \def\url#1{\texttt{#1}}\fi
\expandafter\ifx\csname urlprefix\endcsname\relax\def\urlprefix{URL }\fi
\providecommand{\bibinfo}[2]{#2}
\providecommand{\eprint}[2][]{\url{#2}}

\bibitem[{\citenamefont{Blaizot}(1980)}]{Blaizot:1980tw}
\bibinfo{author}{\bibfnamefont{J.~P.} \bibnamefont{Blaizot}},
  \bibinfo{journal}{Phys. Rept.} \textbf{\bibinfo{volume}{64}},
  \bibinfo{pages}{171} (\bibinfo{year}{1980}).

\bibitem[{\citenamefont{Blaizot et~al.}(1995)\citenamefont{Blaizot, Berger,
  Decharg\'e, and Girod}}]{Blaizot:1995}
\bibinfo{author}{\bibfnamefont{J.~P.} \bibnamefont{Blaizot}},
  \bibinfo{author}{\bibfnamefont{J.~F.} \bibnamefont{Berger}},
  \bibinfo{author}{\bibfnamefont{J.}~\bibnamefont{Decharg\'e}},
  \bibnamefont{and} \bibinfo{author}{\bibfnamefont{M.}~\bibnamefont{Girod}},
  \bibinfo{journal}{Nucl. Phys.} \textbf{\bibinfo{volume}{A591}},
  \bibinfo{pages}{435} (\bibinfo{year}{1995}).

\bibitem[{\citenamefont{Youngblood et~al.}(1977)\citenamefont{Youngblood,
  Rozsa, Moss, Brown, and Bronson}}]{Youngblood:1977}
\bibinfo{author}{\bibfnamefont{D.~H.} \bibnamefont{Youngblood}},
  \bibinfo{author}{\bibfnamefont{C.~M.} \bibnamefont{Rozsa}},
  \bibinfo{author}{\bibfnamefont{J.~M.} \bibnamefont{Moss}},
  \bibinfo{author}{\bibfnamefont{D.~R.} \bibnamefont{Brown}}, \bibnamefont{and}
  \bibinfo{author}{\bibfnamefont{J.~D.} \bibnamefont{Bronson}},
  \bibinfo{journal}{Phys. Rev. Lett.} \textbf{\bibinfo{volume}{39}},
  \bibinfo{pages}{1188} (\bibinfo{year}{1977}).

\bibitem[{\citenamefont{Youngblood et~al.}(1981)\citenamefont{Youngblood,
  Bogucki, Bronson, Garg, Lui, and Rozsa}}]{Youngblood:1981}
\bibinfo{author}{\bibfnamefont{D.~H.} \bibnamefont{Youngblood}},
  \bibinfo{author}{\bibfnamefont{P.}~\bibnamefont{Bogucki}},
  \bibinfo{author}{\bibfnamefont{J.~D.} \bibnamefont{Bronson}},
  \bibinfo{author}{\bibfnamefont{U.}~\bibnamefont{Garg}},
  \bibinfo{author}{\bibfnamefont{Y.~W.} \bibnamefont{Lui}}, \bibnamefont{and}
  \bibinfo{author}{\bibfnamefont{C.~M.} \bibnamefont{Rozsa}},
  \bibinfo{journal}{Phys. Rev. C} \textbf{\bibinfo{volume}{23}},
  \bibinfo{pages}{1997} (\bibinfo{year}{1981}).

\bibitem[{\citenamefont{Youngblood et~al.}(1999)\citenamefont{Youngblood,
  Clark, and Lui}}]{Youngblood:1999}
\bibinfo{author}{\bibfnamefont{D.~H.} \bibnamefont{Youngblood}},
  \bibinfo{author}{\bibfnamefont{H.~L.} \bibnamefont{Clark}}, \bibnamefont{and}
  \bibinfo{author}{\bibfnamefont{Y.-W.} \bibnamefont{Lui}},
  \bibinfo{journal}{Phys. Rev. Lett.} \textbf{\bibinfo{volume}{82}},
  \bibinfo{pages}{691} (\bibinfo{year}{1999}).

\bibitem[{\citenamefont{Pearson}(1991)}]{Pearson:1991}
\bibinfo{author}{\bibfnamefont{J.~M.} \bibnamefont{Pearson}},
  \bibinfo{journal}{Phys. Lett.} \textbf{\bibinfo{volume}{B271}},
  \bibinfo{pages}{12} (\bibinfo{year}{1991}).

\bibitem[{\citenamefont{Shlomo and Youngblood}(1993)}]{Shlomo:1993zz}
\bibinfo{author}{\bibfnamefont{S.}~\bibnamefont{Shlomo}} \bibnamefont{and}
  \bibinfo{author}{\bibfnamefont{D.~H.} \bibnamefont{Youngblood}},
  \bibinfo{journal}{Phys. Rev.} \textbf{\bibinfo{volume}{C47}},
  \bibinfo{pages}{529} (\bibinfo{year}{1993}).

\bibitem[{\citenamefont{Col\`o et~al.}(1992)\citenamefont{Col\`o, Bortignon,
  Giai, Bracco, and Broglia}}]{Colo:1992}
\bibinfo{author}{\bibfnamefont{G.}~\bibnamefont{Col\`o}},
  \bibinfo{author}{\bibfnamefont{P.~F.} \bibnamefont{Bortignon}},
  \bibinfo{author}{\bibfnamefont{N.~V.} \bibnamefont{Giai}},
  \bibinfo{author}{\bibfnamefont{A.}~\bibnamefont{Bracco}}, \bibnamefont{and}
  \bibinfo{author}{\bibfnamefont{R.~A.} \bibnamefont{Broglia}},
  \bibinfo{journal}{Physics Letters B} \textbf{\bibinfo{volume}{276}},
  \bibinfo{pages}{279 } (\bibinfo{year}{1992}), ISSN \bibinfo{issn}{0370-2693}.

\bibitem[{\citenamefont{Hamamoto et~al.}(1997)\citenamefont{Hamamoto, Sagawa,
  and Zhang}}]{Hamamoto:1997b}
\bibinfo{author}{\bibfnamefont{I.}~\bibnamefont{Hamamoto}},
  \bibinfo{author}{\bibfnamefont{H.}~\bibnamefont{Sagawa}}, \bibnamefont{and}
  \bibinfo{author}{\bibfnamefont{X.~Z.} \bibnamefont{Zhang}},
  \bibinfo{journal}{Phys. Rev.} \textbf{\bibinfo{volume}{C56}},
  \bibinfo{pages}{3121} (\bibinfo{year}{1997}).

\bibitem[{\citenamefont{Lalazissis et~al.}(1997)\citenamefont{Lalazissis,
  Konig, and Ring}}]{Lalazissis:1996rd}
\bibinfo{author}{\bibfnamefont{G.~A.} \bibnamefont{Lalazissis}},
  \bibinfo{author}{\bibfnamefont{J.}~\bibnamefont{Konig}}, \bibnamefont{and}
  \bibinfo{author}{\bibfnamefont{P.}~\bibnamefont{Ring}},
  \bibinfo{journal}{Phys. Rev.} \textbf{\bibinfo{volume}{C55}},
  \bibinfo{pages}{540} (\bibinfo{year}{1997}), \eprint{nucl-th/9607039}.

\bibitem[{\citenamefont{Vretenar et~al.}(2001)\citenamefont{Vretenar, Paar,
  Ring, and Lalazissis}}]{Vretenar:2001hs}
\bibinfo{author}{\bibfnamefont{D.}~\bibnamefont{Vretenar}},
  \bibinfo{author}{\bibfnamefont{N.}~\bibnamefont{Paar}},
  \bibinfo{author}{\bibfnamefont{P.}~\bibnamefont{Ring}}, \bibnamefont{and}
  \bibinfo{author}{\bibfnamefont{G.~A.} \bibnamefont{Lalazissis}},
  \bibinfo{journal}{Nucl. Phys.} \textbf{\bibinfo{volume}{A692}},
  \bibinfo{pages}{496} (\bibinfo{year}{2001}), \eprint{nucl-th/0101063}.

\bibitem[{\citenamefont{Piekarewicz}(2002)}]{Piekarewicz:2002jd}
\bibinfo{author}{\bibfnamefont{J.}~\bibnamefont{Piekarewicz}},
  \bibinfo{journal}{Phys. Rev.} \textbf{\bibinfo{volume}{C66}},
  \bibinfo{pages}{034305} (\bibinfo{year}{2002}).

\bibitem[{\citenamefont{Piekarewicz}(2004)}]{Piekarewicz:2003br}
\bibinfo{author}{\bibfnamefont{J.}~\bibnamefont{Piekarewicz}},
  \bibinfo{journal}{Phys. Rev.} \textbf{\bibinfo{volume}{C69}},
  \bibinfo{pages}{041301} (\bibinfo{year}{2004}), \eprint{nucl-th/0312020}.

\bibitem[{\citenamefont{Vretenar et~al.}(2003)\citenamefont{Vretenar, Niksic,
  and Ring}}]{Vretenar:2003qm}
\bibinfo{author}{\bibfnamefont{D.}~\bibnamefont{Vretenar}},
  \bibinfo{author}{\bibfnamefont{T.}~\bibnamefont{Niksic}}, \bibnamefont{and}
  \bibinfo{author}{\bibfnamefont{P.}~\bibnamefont{Ring}},
  \bibinfo{journal}{Phys. Rev.} \textbf{\bibinfo{volume}{C68}},
  \bibinfo{pages}{024310} (\bibinfo{year}{2003}), \eprint{nucl-th/0302070}.

\bibitem[{\citenamefont{Agrawal et~al.}(2003)\citenamefont{Agrawal, Shlomo, and
  Au}}]{Agrawal:2003xb}
\bibinfo{author}{\bibfnamefont{B.~K.} \bibnamefont{Agrawal}},
  \bibinfo{author}{\bibfnamefont{S.}~\bibnamefont{Shlomo}}, \bibnamefont{and}
  \bibinfo{author}{\bibfnamefont{V.~K.} \bibnamefont{Au}},
  \bibinfo{journal}{Phys. Rev.} \textbf{\bibinfo{volume}{C68}},
  \bibinfo{pages}{031304} (\bibinfo{year}{2003}), \eprint{nucl-th/0308042}.

\bibitem[{\citenamefont{Col\`o et~al.}(2004)\citenamefont{Col\`o, Van~Giai,
  Meyer, Bennaceur, and Bonche}}]{Colo:2004mj}
\bibinfo{author}{\bibfnamefont{G.}~\bibnamefont{Col\`o}},
  \bibinfo{author}{\bibfnamefont{N.}~\bibnamefont{Van~Giai}},
  \bibinfo{author}{\bibfnamefont{J.}~\bibnamefont{Meyer}},
  \bibinfo{author}{\bibfnamefont{K.}~\bibnamefont{Bennaceur}},
  \bibnamefont{and} \bibinfo{author}{\bibfnamefont{P.}~\bibnamefont{Bonche}},
  \bibinfo{journal}{Phys. Rev.} \textbf{\bibinfo{volume}{C70}},
  \bibinfo{pages}{024307} (\bibinfo{year}{2004}), \eprint{nucl-th/0403086}.

\bibitem[{\citenamefont{Todd-Rutel and Piekarewicz}(2005)}]{Todd-Rutel:2005fa}
\bibinfo{author}{\bibfnamefont{B.~G.} \bibnamefont{Todd-Rutel}}
  \bibnamefont{and}
  \bibinfo{author}{\bibfnamefont{J.}~\bibnamefont{Piekarewicz}},
  \bibinfo{journal}{Phys. Rev. Lett} \textbf{\bibinfo{volume}{95}},
  \bibinfo{pages}{122501} (\bibinfo{year}{2005}), \eprint{nucl-th/0504034}.

\bibitem[{\citenamefont{Agrawal et~al.}(2005)\citenamefont{Agrawal, Shlomo, and
  Au}}]{Agrawal:2005ix}
\bibinfo{author}{\bibfnamefont{B.~K.} \bibnamefont{Agrawal}},
  \bibinfo{author}{\bibfnamefont{S.}~\bibnamefont{Shlomo}}, \bibnamefont{and}
  \bibinfo{author}{\bibfnamefont{V.~K.} \bibnamefont{Au}},
  \bibinfo{journal}{Phys. Rev.} \textbf{\bibinfo{volume}{C72}},
  \bibinfo{pages}{0143310} (\bibinfo{year}{2005}), \eprint{nucl-th/0505071}.

\bibitem[{\citenamefont{Lalazissis et~al.}(2005)\citenamefont{Lalazissis,
  Niksic, Vretenar, and Ring}}]{Lalazissis:2005de}
\bibinfo{author}{\bibfnamefont{G.~A.} \bibnamefont{Lalazissis}},
  \bibinfo{author}{\bibfnamefont{T.}~\bibnamefont{Niksic}},
  \bibinfo{author}{\bibfnamefont{D.}~\bibnamefont{Vretenar}}, \bibnamefont{and}
  \bibinfo{author}{\bibfnamefont{P.}~\bibnamefont{Ring}},
  \bibinfo{journal}{Phys. Rev.} \textbf{\bibinfo{volume}{C71}},
  \bibinfo{pages}{024312} (\bibinfo{year}{2005}).

\bibitem[{\citenamefont{Garg et~al.}(2007)}]{Garg:2006vc}
\bibinfo{author}{\bibfnamefont{U.}~\bibnamefont{Garg}} \bibnamefont{et~al.},
  \bibinfo{journal}{Nucl. Phys.} \textbf{\bibinfo{volume}{A788}},
  \bibinfo{pages}{36} (\bibinfo{year}{2007}), \eprint{nucl-ex/0608007}.

\bibitem[{\citenamefont{Piekarewicz and Centelles}(2009)}]{Piekarewicz:2008nh}
\bibinfo{author}{\bibfnamefont{J.}~\bibnamefont{Piekarewicz}} \bibnamefont{and}
  \bibinfo{author}{\bibfnamefont{M.}~\bibnamefont{Centelles}},
  \bibinfo{journal}{Phys. Rev.} \textbf{\bibinfo{volume}{C79}},
  \bibinfo{pages}{054311} (\bibinfo{year}{2009}), \eprint{0812.4499}.

\bibitem[{\citenamefont{Li et~al.}(2007)}]{Li:2007bp}
\bibinfo{author}{\bibfnamefont{T.}~\bibnamefont{Li}} \bibnamefont{et~al.},
  \bibinfo{journal}{Phys. Rev. Lett.} \textbf{\bibinfo{volume}{99}},
  \bibinfo{pages}{162503} (\bibinfo{year}{2007}), \eprint{0709.0567}.

\bibitem[{\citenamefont{Horowitz et~al.}(2001)\citenamefont{Horowitz, Pollock,
  Souder, and Michaels}}]{Horowitz:1999fk}
\bibinfo{author}{\bibfnamefont{C.~J.} \bibnamefont{Horowitz}},
  \bibinfo{author}{\bibfnamefont{S.~J.} \bibnamefont{Pollock}},
  \bibinfo{author}{\bibfnamefont{P.~A.} \bibnamefont{Souder}},
  \bibnamefont{and} \bibinfo{author}{\bibfnamefont{R.}~\bibnamefont{Michaels}},
  \bibinfo{journal}{Phys. Rev.} \textbf{\bibinfo{volume}{C63}},
  \bibinfo{pages}{025501} (\bibinfo{year}{2001}), \eprint{nucl-th/9912038}.

\bibitem[{\citenamefont{Kumar et~al.}(2005)\citenamefont{Kumar, Michaels,
  Souder, and Urciuoli}}]{Michaels:2005}
\bibinfo{author}{\bibfnamefont{K.}~\bibnamefont{Kumar}},
  \bibinfo{author}{\bibfnamefont{R.}~\bibnamefont{Michaels}},
  \bibinfo{author}{\bibfnamefont{P.~A.} \bibnamefont{Souder}},
  \bibnamefont{and} \bibinfo{author}{\bibfnamefont{G.~M.}
  \bibnamefont{Urciuoli}} (\bibinfo{year}{2005}),
  \urlprefix\url{http://hallaweb.jlab.org/parity/prex}.

\bibitem[{\citenamefont{Todd and Piekarewicz}(2003)}]{Todd:2003xs}
\bibinfo{author}{\bibfnamefont{B.~G.} \bibnamefont{Todd}} \bibnamefont{and}
  \bibinfo{author}{\bibfnamefont{J.}~\bibnamefont{Piekarewicz}},
  \bibinfo{journal}{Phys. Rev.} \textbf{\bibinfo{volume}{C67}},
  \bibinfo{pages}{044317} (\bibinfo{year}{2003}), \eprint{nucl-th/0301092}.

\bibitem[{\citenamefont{Danielewicz et~al.}(2002)\citenamefont{Danielewicz,
  Lacey, and Lynch}}]{Danielewicz:2002pu}
\bibinfo{author}{\bibfnamefont{P.}~\bibnamefont{Danielewicz}},
  \bibinfo{author}{\bibfnamefont{R.}~\bibnamefont{Lacey}}, \bibnamefont{and}
  \bibinfo{author}{\bibfnamefont{W.~G.} \bibnamefont{Lynch}},
  \bibinfo{journal}{Science} \textbf{\bibinfo{volume}{298}},
  \bibinfo{pages}{1592} (\bibinfo{year}{2002}), \eprint{nucl-th/0208016}.

\bibitem[{\citenamefont{Tsang et~al.}(2004)}]{Tsang:2004}
\bibinfo{author}{\bibfnamefont{M.~B.} \bibnamefont{Tsang}}
  \bibnamefont{et~al.}, \bibinfo{journal}{Phys. Rev. Lett.}
  \textbf{\bibinfo{volume}{92}}, \bibinfo{pages}{062701}
  (\bibinfo{year}{2004}).

\bibitem[{\citenamefont{Chen et~al.}(2005)\citenamefont{Chen, Ko, and
  Li}}]{Chen:2004si}
\bibinfo{author}{\bibfnamefont{L.-W.} \bibnamefont{Chen}},
  \bibinfo{author}{\bibfnamefont{C.~M.} \bibnamefont{Ko}}, \bibnamefont{and}
  \bibinfo{author}{\bibfnamefont{B.-A.} \bibnamefont{Li}},
  \bibinfo{journal}{Phys. Rev. Lett.} \textbf{\bibinfo{volume}{94}},
  \bibinfo{pages}{032701} (\bibinfo{year}{2005}), \eprint{nucl-th/0407032}.

\bibitem[{\citenamefont{Li and Steiner}(2006)}]{Li:2005sr}
\bibinfo{author}{\bibfnamefont{B.-A.} \bibnamefont{Li}} \bibnamefont{and}
  \bibinfo{author}{\bibfnamefont{A.~W.} \bibnamefont{Steiner}},
  \bibinfo{journal}{Phys. Lett.} \textbf{\bibinfo{volume}{B642}},
  \bibinfo{pages}{436} (\bibinfo{year}{2006}), \eprint{nucl-th/0511064}.

\bibitem[{\citenamefont{Shetty et~al.}(2007)\citenamefont{Shetty, Yennello, and
  Souliotis}}]{Shetty:2005qp}
\bibinfo{author}{\bibfnamefont{D.~V.} \bibnamefont{Shetty}},
  \bibinfo{author}{\bibfnamefont{S.~J.} \bibnamefont{Yennello}},
  \bibnamefont{and} \bibinfo{author}{\bibfnamefont{G.~A.}
  \bibnamefont{Souliotis}}, \bibinfo{journal}{Phys. Rev.}
  \textbf{\bibinfo{volume}{C75}}, \bibinfo{pages}{034602}
  (\bibinfo{year}{2007}), \eprint{nucl-ex/0505011}.

\bibitem[{\citenamefont{Sil et~al.}(2005)\citenamefont{Sil, Centelles, Vinas,
  and Piekarewicz}}]{Sil:2005tg}
\bibinfo{author}{\bibfnamefont{T.}~\bibnamefont{Sil}},
  \bibinfo{author}{\bibfnamefont{M.}~\bibnamefont{Centelles}},
  \bibinfo{author}{\bibfnamefont{X.}~\bibnamefont{Vinas}}, \bibnamefont{and}
  \bibinfo{author}{\bibfnamefont{J.}~\bibnamefont{Piekarewicz}},
  \bibinfo{journal}{Phys. Rev.} \textbf{\bibinfo{volume}{C71}},
  \bibinfo{pages}{045502} (\bibinfo{year}{2005}), \eprint{nucl-th/0501014}.

\bibitem[{\citenamefont{Piekarewicz}(2007{\natexlab{a}})}]{Piekarewicz:2006vp}
\bibinfo{author}{\bibfnamefont{J.}~\bibnamefont{Piekarewicz}},
  \bibinfo{journal}{Eur. Phys. J.} \textbf{\bibinfo{volume}{A32}},
  \bibinfo{pages}{537} (\bibinfo{year}{2007}{\natexlab{a}}),
  \eprint{nucl-th/0608024}.

\bibitem[{\citenamefont{Horowitz and
  Piekarewicz}(2001{\natexlab{a}})}]{Horowitz:2000xj}
\bibinfo{author}{\bibfnamefont{C.~J.} \bibnamefont{Horowitz}} \bibnamefont{and}
  \bibinfo{author}{\bibfnamefont{J.}~\bibnamefont{Piekarewicz}},
  \bibinfo{journal}{Phys. Rev. Lett.} \textbf{\bibinfo{volume}{86}},
  \bibinfo{pages}{5647} (\bibinfo{year}{2001}{\natexlab{a}}),
  \eprint{astro-ph/0010227}.

\bibitem[{\citenamefont{Horowitz and
  Piekarewicz}(2001{\natexlab{b}})}]{Horowitz:2001ya}
\bibinfo{author}{\bibfnamefont{C.~J.} \bibnamefont{Horowitz}} \bibnamefont{and}
  \bibinfo{author}{\bibfnamefont{J.}~\bibnamefont{Piekarewicz}},
  \bibinfo{journal}{Phys. Rev.} \textbf{\bibinfo{volume}{C64}},
  \bibinfo{pages}{062802} (\bibinfo{year}{2001}{\natexlab{b}}),
  \eprint{nucl-th/0108036}.

\bibitem[{\citenamefont{Horowitz and Piekarewicz}(2002)}]{Horowitz:2002mb}
\bibinfo{author}{\bibfnamefont{C.~J.} \bibnamefont{Horowitz}} \bibnamefont{and}
  \bibinfo{author}{\bibfnamefont{J.}~\bibnamefont{Piekarewicz}},
  \bibinfo{journal}{Phys. Rev.} \textbf{\bibinfo{volume}{C66}},
  \bibinfo{pages}{055803} (\bibinfo{year}{2002}), \eprint{nucl-th/0207067}.

\bibitem[{\citenamefont{Carriere et~al.}(2003)\citenamefont{Carriere, Horowitz,
  and Piekarewicz}}]{Carriere:2002bx}
\bibinfo{author}{\bibfnamefont{J.}~\bibnamefont{Carriere}},
  \bibinfo{author}{\bibfnamefont{C.~J.} \bibnamefont{Horowitz}},
  \bibnamefont{and}
  \bibinfo{author}{\bibfnamefont{J.}~\bibnamefont{Piekarewicz}},
  \bibinfo{journal}{Astrophys. J.} \textbf{\bibinfo{volume}{593}},
  \bibinfo{pages}{463} (\bibinfo{year}{2003}), \eprint{nucl-th/0211015}.

\bibitem[{\citenamefont{Buras et~al.}(2003)\citenamefont{Buras, Rampp, Janka,
  and Kifonidis}}]{Buras:2003sn}
\bibinfo{author}{\bibfnamefont{R.}~\bibnamefont{Buras}},
  \bibinfo{author}{\bibfnamefont{M.}~\bibnamefont{Rampp}},
  \bibinfo{author}{\bibfnamefont{H.~T.} \bibnamefont{Janka}}, \bibnamefont{and}
  \bibinfo{author}{\bibfnamefont{K.}~\bibnamefont{Kifonidis}},
  \bibinfo{journal}{Phys. Rev. Lett.} \textbf{\bibinfo{volume}{90}},
  \bibinfo{pages}{241101} (\bibinfo{year}{2003}), \eprint{astro-ph/0303171}.

\bibitem[{\citenamefont{Lattimer and Prakash}(2004)}]{Lattimer:2004pg}
\bibinfo{author}{\bibfnamefont{J.~M.} \bibnamefont{Lattimer}} \bibnamefont{and}
  \bibinfo{author}{\bibfnamefont{M.}~\bibnamefont{Prakash}},
  \bibinfo{journal}{Science} \textbf{\bibinfo{volume}{304}},
  \bibinfo{pages}{536} (\bibinfo{year}{2004}), \eprint{astro-ph/0405262}.

\bibitem[{\citenamefont{Lattimer and Prakash}(2007)}]{Lattimer:2006xb}
\bibinfo{author}{\bibfnamefont{J.~M.} \bibnamefont{Lattimer}} \bibnamefont{and}
  \bibinfo{author}{\bibfnamefont{M.}~\bibnamefont{Prakash}},
  \bibinfo{journal}{Phys. Rept.} \textbf{\bibinfo{volume}{442}},
  \bibinfo{pages}{109} (\bibinfo{year}{2007}), \eprint{astro-ph/0612440}.

\bibitem[{\citenamefont{Steiner et~al.}(2005)\citenamefont{Steiner, Prakash,
  Lattimer, and Ellis}}]{Steiner:2004fi}
\bibinfo{author}{\bibfnamefont{A.~W.} \bibnamefont{Steiner}},
  \bibinfo{author}{\bibfnamefont{M.}~\bibnamefont{Prakash}},
  \bibinfo{author}{\bibfnamefont{J.~M.} \bibnamefont{Lattimer}},
  \bibnamefont{and} \bibinfo{author}{\bibfnamefont{P.~J.} \bibnamefont{Ellis}},
  \bibinfo{journal}{Phys. Rept.} \textbf{\bibinfo{volume}{411}},
  \bibinfo{pages}{325} (\bibinfo{year}{2005}), \eprint{nucl-th/0410066}.

\bibitem[{\citenamefont{Piekarewicz}(2007{\natexlab{b}})}]{Piekarewicz:2007us}
\bibinfo{author}{\bibfnamefont{J.}~\bibnamefont{Piekarewicz}},
  \bibinfo{journal}{Phys. Rev.} \textbf{\bibinfo{volume}{C76}},
  \bibinfo{pages}{031301} (\bibinfo{year}{2007}{\natexlab{b}}),
  \eprint{0705.1491}.

\bibitem[{\citenamefont{Sagawa et~al.}(2007)\citenamefont{Sagawa, Yoshida,
  Zeng, Gu, and Zhang}}]{Sagawa:2007sp}
\bibinfo{author}{\bibfnamefont{H.}~\bibnamefont{Sagawa}},
  \bibinfo{author}{\bibfnamefont{S.}~\bibnamefont{Yoshida}},
  \bibinfo{author}{\bibfnamefont{G.-M.} \bibnamefont{Zeng}},
  \bibinfo{author}{\bibfnamefont{J.-Z.} \bibnamefont{Gu}}, \bibnamefont{and}
  \bibinfo{author}{\bibfnamefont{X.-Z.} \bibnamefont{Zhang}},
  \bibinfo{journal}{Phys. Rev.} \textbf{\bibinfo{volume}{C76}},
  \bibinfo{pages}{034327} (\bibinfo{year}{2007}), \eprint{0706.0966}.

\bibitem[{\citenamefont{Tselyaev et~al.}(2009)\citenamefont{Tselyaev, Speth,
  Krewald, Litvinova, Kamerdzhiev, Lyutorovich, Avdeenkov, and
  Gr\"{u}mmer}}]{Tselyaev:2009}
\bibinfo{author}{\bibfnamefont{V.}~\bibnamefont{Tselyaev}},
  \bibinfo{author}{\bibfnamefont{J.}~\bibnamefont{Speth}},
  \bibinfo{author}{\bibfnamefont{S.}~\bibnamefont{Krewald}},
  \bibinfo{author}{\bibfnamefont{E.}~\bibnamefont{Litvinova}},
  \bibinfo{author}{\bibfnamefont{S.}~\bibnamefont{Kamerdzhiev}},
  \bibinfo{author}{\bibfnamefont{N.}~\bibnamefont{Lyutorovich}},
  \bibinfo{author}{\bibfnamefont{A.}~\bibnamefont{Avdeenkov}},
  \bibnamefont{and}
  \bibinfo{author}{\bibfnamefont{F.}~\bibnamefont{Gr\"{u}mmer}},
  \bibinfo{journal}{Phys. Rev.} \textbf{\bibinfo{volume}{C79}},
  \bibinfo{eid}{034309} (\bibinfo{year}{2009}).

\bibitem[{\citenamefont{Piekarewicz}(2000)}]{Piekarewicz:2000nm}
\bibinfo{author}{\bibfnamefont{J.}~\bibnamefont{Piekarewicz}},
  \bibinfo{journal}{Phys. Rev.} \textbf{\bibinfo{volume}{C62}},
  \bibinfo{pages}{051304} (\bibinfo{year}{2000}), \eprint{nucl-th/0003029}.

\bibitem[{\citenamefont{Piekarewicz}(2001)}]{Piekarewicz:2001nm}
\bibinfo{author}{\bibfnamefont{J.}~\bibnamefont{Piekarewicz}},
  \bibinfo{journal}{Phys. Rev.} \textbf{\bibinfo{volume}{C64}},
  \bibinfo{pages}{024307} (\bibinfo{year}{2001}), \eprint{nucl-th/0103016}.

\bibitem[{\citenamefont{Bohr and Mottelson}(1998)}]{BohrII:1998}
\bibinfo{author}{\bibfnamefont{A.}~\bibnamefont{Bohr}} \bibnamefont{and}
  \bibinfo{author}{\bibfnamefont{B.~R.} \bibnamefont{Mottelson}},
  \emph{\bibinfo{title}{Nuclear Structure}} (\bibinfo{publisher}{World
  Scientific Publishing Company, New Jersey}, \bibinfo{year}{1998}), vol.
  \bibinfo{volume}{II: Nuclear Deformations}.

\bibitem[{\citenamefont{Harakeh and van~der Woude}(2001)}]{Harakeh:2001}
\bibinfo{author}{\bibfnamefont{M.~N.} \bibnamefont{Harakeh}} \bibnamefont{and}
  \bibinfo{author}{\bibfnamefont{A.}~\bibnamefont{van~der Woude}},
  \emph{\bibinfo{title}{Giant Resonances-Fundamental High-frequency Modes of
  Nuclear Excitation}} (\bibinfo{publisher}{Clarendon, Oxford},
  \bibinfo{year}{2001}).

\bibitem[{\citenamefont{Lalazissis et~al.}(1999)\citenamefont{Lalazissis,
  Raman, and Ring}}]{Lalazissis:1999}
\bibinfo{author}{\bibfnamefont{G.~A.} \bibnamefont{Lalazissis}},
  \bibinfo{author}{\bibfnamefont{S.}~\bibnamefont{Raman}}, \bibnamefont{and}
  \bibinfo{author}{\bibfnamefont{P.}~\bibnamefont{Ring}}, \bibinfo{journal}{At.
  Data Nucl. Data Tables} \textbf{\bibinfo{volume}{71}}, \bibinfo{pages}{1}
  (\bibinfo{year}{1999}).

\bibitem[{\citenamefont{Lui et~al.}(2004)\citenamefont{Lui, Youngblood,
  Tokimoto, Clark, and John}}]{Lui:2004wm}
\bibinfo{author}{\bibfnamefont{Y.~W.} \bibnamefont{Lui}},
  \bibinfo{author}{\bibfnamefont{D.~H.} \bibnamefont{Youngblood}},
  \bibinfo{author}{\bibfnamefont{Y.}~\bibnamefont{Tokimoto}},
  \bibinfo{author}{\bibfnamefont{H.~L.} \bibnamefont{Clark}}, \bibnamefont{and}
  \bibinfo{author}{\bibfnamefont{B.}~\bibnamefont{John}},
  \bibinfo{journal}{Phys. Rev.} \textbf{\bibinfo{volume}{C70}},
  \bibinfo{pages}{014307} (\bibinfo{year}{2004}).

\bibitem[{\citenamefont{Li et~al.}(2008)\citenamefont{Li, Colo, and
  Meng}}]{Li:2008hx}
\bibinfo{author}{\bibfnamefont{J.}~\bibnamefont{Li}},
  \bibinfo{author}{\bibfnamefont{G.}~\bibnamefont{Colo}}, \bibnamefont{and}
  \bibinfo{author}{\bibfnamefont{J.}~\bibnamefont{Meng}},
  \bibinfo{journal}{Phys. Rev.} \textbf{\bibinfo{volume}{C78}},
  \bibinfo{pages}{064304} (\bibinfo{year}{2008}), \eprint{0811.4091}.

\bibitem[{\citenamefont{Khan}(2009)}]{Khan:2009xq}
\bibinfo{author}{\bibfnamefont{E.}~\bibnamefont{Khan}}, \bibinfo{journal}{Phys.
  Rev.} \textbf{\bibinfo{volume}{C80}}, \bibinfo{pages}{011307}
  (\bibinfo{year}{2009}), \eprint{0905.3335}.

\bibitem[{\citenamefont{Friedman and Pandharipande}(1981)}]{Friedman:1981qw}
\bibinfo{author}{\bibfnamefont{B.}~\bibnamefont{Friedman}} \bibnamefont{and}
  \bibinfo{author}{\bibfnamefont{V.~R.} \bibnamefont{Pandharipande}},
  \bibinfo{journal}{Nucl. Phys.} \textbf{\bibinfo{volume}{A361}},
  \bibinfo{pages}{502} (\bibinfo{year}{1981}).

\bibitem[{\citenamefont{Schwenk and Pethick}(2005)}]{Schwenk:2005ka}
\bibinfo{author}{\bibfnamefont{A.}~\bibnamefont{Schwenk}} \bibnamefont{and}
  \bibinfo{author}{\bibfnamefont{C.~J.} \bibnamefont{Pethick}},
  \bibinfo{journal}{Phys. Rev. Lett.} \textbf{\bibinfo{volume}{95}},
  \bibinfo{pages}{160401} (\bibinfo{year}{2005}), \eprint{nucl-th/0506042}.

\bibitem[{\citenamefont{Hebeler and Schwenk}(2009)}]{Hebeler:2009iv}
\bibinfo{author}{\bibfnamefont{K.}~\bibnamefont{Hebeler}} \bibnamefont{and}
  \bibinfo{author}{\bibfnamefont{A.}~\bibnamefont{Schwenk}}
  (\bibinfo{year}{2009}), \eprint{0911.0483}.

\bibitem[{\citenamefont{Gandolfi et~al.}(2008)\citenamefont{Gandolfi,
  Illarionov, Fantoni, Pederiva, and Schmidt}}]{Gandolfi:2008id}
\bibinfo{author}{\bibfnamefont{S.}~\bibnamefont{Gandolfi}},
  \bibinfo{author}{\bibfnamefont{A.~Y.} \bibnamefont{Illarionov}},
  \bibinfo{author}{\bibfnamefont{S.}~\bibnamefont{Fantoni}},
  \bibinfo{author}{\bibfnamefont{F.}~\bibnamefont{Pederiva}}, \bibnamefont{and}
  \bibinfo{author}{\bibfnamefont{K.~E.} \bibnamefont{Schmidt}},
  \bibinfo{journal}{Phys. Rev. Lett.} \textbf{\bibinfo{volume}{101}},
  \bibinfo{pages}{132501} (\bibinfo{year}{2008}), \eprint{0805.2513}.

\bibitem[{\citenamefont{Gezerlis and Carlson}(2009)}]{Gezerlis:2009iw}
\bibinfo{author}{\bibfnamefont{A.}~\bibnamefont{Gezerlis}} \bibnamefont{and}
  \bibinfo{author}{\bibfnamefont{J.}~\bibnamefont{Carlson}}
  (\bibinfo{year}{2009}), \eprint{0911.3907}.

\bibitem[{\citenamefont{Baker}(1999)}]{Baker:1999dg}
\bibinfo{author}{\bibfnamefont{G.~A.} \bibnamefont{Baker}},
  \bibinfo{journal}{Phys. Rev.} \textbf{\bibinfo{volume}{C60}},
  \bibinfo{pages}{054311} (\bibinfo{year}{1999}).

\bibitem[{\citenamefont{Heiselberg}(2002)}]{Heiselberg:2000bm}
\bibinfo{author}{\bibfnamefont{H.}~\bibnamefont{Heiselberg}},
  \bibinfo{journal}{Phys. Rev.} \textbf{\bibinfo{volume}{A63}},
  \bibinfo{pages}{043606} (\bibinfo{year}{2002}), \eprint{cond-mat/0002056}.

\bibitem[{\citenamefont{Carlson et~al.}(2003)\citenamefont{Carlson, Chang,
  Pandharipande, and Schmidt}}]{Carlson:2003}
\bibinfo{author}{\bibfnamefont{J.}~\bibnamefont{Carlson}},
  \bibinfo{author}{\bibfnamefont{S.-Y.} \bibnamefont{Chang}},
  \bibinfo{author}{\bibfnamefont{V.~R.} \bibnamefont{Pandharipande}},
  \bibnamefont{and} \bibinfo{author}{\bibfnamefont{K.~E.}
  \bibnamefont{Schmidt}}, \bibinfo{journal}{Phys. Rev. Lett.}
  \textbf{\bibinfo{volume}{91}}, \bibinfo{pages}{050401}
  (\bibinfo{year}{2003}).

\bibitem[{\citenamefont{Nishida and Son}(2006)}]{Nishida:2006br}
\bibinfo{author}{\bibfnamefont{Y.}~\bibnamefont{Nishida}} \bibnamefont{and}
  \bibinfo{author}{\bibfnamefont{D.~T.} \bibnamefont{Son}},
  \bibinfo{journal}{Phys. Rev. Lett.} \textbf{\bibinfo{volume}{97}},
  \bibinfo{pages}{050403} (\bibinfo{year}{2006}), \eprint{cond-mat/0604500}.

\bibitem[{\citenamefont{Chen et~al.}(2007)\citenamefont{Chen, Ko, and
  Li}}]{Chen:2007ih}
\bibinfo{author}{\bibfnamefont{L.-W.} \bibnamefont{Chen}},
  \bibinfo{author}{\bibfnamefont{C.~M.} \bibnamefont{Ko}}, \bibnamefont{and}
  \bibinfo{author}{\bibfnamefont{B.-A.} \bibnamefont{Li}},
  \bibinfo{journal}{Phys. Rev.} \textbf{\bibinfo{volume}{C76}},
  \bibinfo{pages}{054316} (\bibinfo{year}{2007}), \eprint{0709.0900}.

\bibitem[{\citenamefont{Chen et~al.}(2009)}]{Chen:2009wv}
\bibinfo{author}{\bibfnamefont{L.-W.} \bibnamefont{Chen}} \bibnamefont{et~al.},
  \bibinfo{journal}{Phys. Rev.} \textbf{\bibinfo{volume}{C80}},
  \bibinfo{pages}{014322} (\bibinfo{year}{2009}), \eprint{0905.4323}.

\end{thebibliography}

\end{document}